\documentclass[final,5p,times,twocolumn,authoryear]{elsarticle}

\usepackage{amssymb}
\usepackage{lipsum}
\usepackage{amsthm}
\usepackage{subcaption}
\usepackage{amsmath}
\usepackage{xcolor}
\usepackage{hyperref}
\usepackage{tikz}
\usetikzlibrary{decorations.pathmorphing,arrows.meta,bending,positioning}

\theoremstyle{definition}
\newtheorem{definition}{Definition}[section]

\newtheorem{theorem}[definition]{Theorem}

\journal{Nuclear Physics B}

\begin{document}

\begin{frontmatter}

\title{Penrose's singularity theorem and the Kerr space-time}

\author[aff1]{Jonathan Brook}
\ead{jonathan.brook@pg.canterbury.ac.nz}
\author[aff1]{Chris Stevens}
\ead{chris.stevens@canterbury.ac.nz}
\affiliation[aff1]{organization={School of Mathematics and Statistics, University of Canterbury},
            city={Christchurch},
            postcode={8041}, 
            country={New Zealand}}

\begin{abstract}
In this short paper, Penrose's famous singularity theorem is applied to the Kerr space-time. In the case of the maximally extended space-time, the assumptions of Penrose's singularity theorem are not satisfied as the space-time is not globally hyperbolic. In the case of the unextended space-time -- defined up to some radius between the inner and outer event horizons -- the assumptions of the theorem hold, but scalar curvature invariants remain finite everywhere. Calculations are done in detail showcasing the validity of the theorem, and misconceptions regarding the characterization of physical singularities by incomplete null geodesics are discussed.
\end{abstract}


\end{frontmatter}

\section{Introduction}
\label{sec:1}

In 1939, Oppenheimer and Snyder showed that a collapsing spherical dust cloud creates a physical (curvature) singularity \citep{oppenheimer1939continued}. However, the question of whether physical singularity formation extends to generic, non-symmetric gravitational collapse was still left open -- a much harder problem to tackle. In 1965, Roger Penrose published a letter that put forward five statements regarding space-time that together are inconsistent \citep{penrosesingularitytheorem}, with one of them being a future null completeness condition. Given the issue of rigorously defining what a physical singularity actually is -- with difficulty arising due to the surface not actually lying in the space-time -- this was put forward as a new method of characterization. This makes sense, as future null incompleteness colloquially is an abrupt halt of future directed null geodesics at a finite affine time. This is defined more rigorously later. Furthermore, the notion of a \emph{trapped surface} was introduced for the first time, with its existence being a critical condition for null incompleteness. This style of singularity characterization led to the birth of the so-called singularity theorems and opened a completely new avenue of mathematical general relativity (see \citep{senovilla2011singularity} for an overview). Penrose most deservingly won half of the 2020 Nobel prize in physics for this contribution.

Despite their success, there are subtleties to the singularity theorems that are often misconstrued (for a very nice, detailed account see \citep{senovilla2022critical}). The first is that Penrose's theorem does not actually say anything about the process of gravitational collapse itself, rather only what happens when a trapped surface already exists. The theorem says, in very generic circumstances, what happens \emph{inside} a black hole. This is extremely important in and of itself, but should not be confused with the process of gravitational collapse and trapped surface \emph{formation}.

The second misconception, which is the focus of this letter, is that the characterization of a physical singularity by the existence of incomplete null geodesics is not always a good one \citep{scott2021singularity}. In many cases, this just predicts that the space-time may be \emph{extended} \footnote{This is the gist of Roy Kerr's argument in \citep{kerr2023blackholessingularities}.}. For example, the Kerr space-time in Kerr-Schild coordinates is \citep{kerr1965new, visser2008kerrspacetimebriefintroduction}
\begin{align}
    \text{d}s^2 
    &= -\text{d}t^2 + \text{d}x^2 + \text{d}y^2 + \text{d}z^2 \nonumber \\
    &\quad+ \frac{2Mr^3}{r^4 + a^2z^2}\left(
    \text{d}t + \frac{r(x\text{d}x + y\text{d}y)}{a^2 + r^2}
    + \frac{a(y\text{d}x - x\text{d}y)}{a^2 + r^2}
    + \frac{z}{r}\text{d}z
    \right)^2, \label{eq:KerrLE}
\end{align}
where $r$ is implicitly defined by
\begin{equation}
    x^2 + y^2 + z^2 = r^2 + a^2\left(1 - \frac{z^2}{r^2}\right) \label{eq:KerrAuxDefs}
\end{equation}
and where $M$ is the mass, $J$ the angular momentum and $a = J/M$. The inner and outer event horizons are located at $r_\pm = M \pm \sqrt{M^2 - a^2}$. It turns out that, with the space-time defined only for $r \geq r_-$, the space-time satisfies all the criteria for Penrose's singularity theorem, which then implies that there exist future-directed null geodesics that are incomplete. However, $r_-$ is not a singularity in the sense that scalar curvature invariants -- scalars formed from curvature quantities such as the Riemann tensor and are independent of the choice of frame, such as the Kretschmann scalar $R_{\alpha\beta\gamma\delta}R^{\alpha\beta\gamma\delta}$ -- diverge. In fact, restricting the space-time to satisfy $r \geq r_0$ for some $ r_- \leq r_0 \leq r_+$, so that the 2-surface $t=t_0,\;r=r_0$ is a trapped surface, would still satisfy Penrose's singularity theorem, but with the metric and all curvature quantities finite at $r_0$. So, there is a subtlety in classifying a physical singularity from the existence of incomplete geodesics. 

The inner horizon $r_-$ is special in that no matter how one extends the space-time past this surface, the extended space-time \emph{will no longer be globally hyperbolic}. For this reason, the inner horizon $r_-$ is a \emph{Cauchy horizon} and is the smallest value of $r$ that one can define the space-time up to before the theorem no longer holds. Further, although there is a unique maximal analytic extension for both the Schwarzschild and Kerr space-times in vacuum through this surface \citep{kruskal1960maximal,boyer1967maximal}, these extensions are not unique if one considers non-vacuum solutions. 

Hence, there are two cases for the Kerr space-time. Firstly, one can consider the space-time defined only for $r \geq r_-$ (or some finite $r=r_0$, which together with a fixed $t$ yields a trapped surface). In this case, Penrose's theorem applies, but the prediction is not a singularity in the sense of unbound scalar curvature invariants. Alternatively, one could extend the space-time beyond $r_-$. In this case, however, the theorem's assumptions are not satisfied and the theorem does not tell us anything, even though the maximal analytic extension contains timelike curvature singularities.

This letter is organized as follows: Sec.~\ref{sec:PenrosesSingThm} outlines Penrose's singularity theorem and discusses its application to the Schwarzschild and Kerr space-times, Sec.~\ref{sec:AppToKerr} gives a fully worked example of the Penrose singularity theorem applied to the Kerr space-time, and Sec.~\ref{sec:discussion} discusses these results and their interpretation.

\section{Penrose's singularity theorem}
\label{sec:PenrosesSingThm}

\subsection{Overview of Penrose's singularity theorem}
\begin{figure}[h!]
    \begin{center}
      \begin{tikzpicture}[very thick, decoration = {bent,amplitude=-5}]
        \draw[dashed] (2,-2) -- (0,0) -- (2,2);
        \draw (2,2) -- (4,0) -- (2,-2);
        \draw[dashed] (-2,-2) -- (0,0) -- (-2,2);
        \draw (-2,-2) -- (-4,0) -- (-2,2);
        \draw[red,decoration={bumps,amplitude=-2},decorate] (-2,2) -- (0, 1.8) -- (2,2);
        \draw[red,decoration={bumps,amplitude=2},decorate] (-2,-2) -- (0, -1.8)-- (2,-2);

        \draw[blue] (-4,0) to[out=15, in=195] (4,0);

        \node at (0,-1) {\large I};
        \node at (-2,0) {\large II};
        \node at (2,0) {\large III};
        \node at (0,1) {\large IV};
      \end{tikzpicture}
    \end{center}
    \caption{The usual Penrose-Carter diagram of a Schwarzschild black hole, with a Cauchy hypersurface drawn in blue across the horizontal.}
    \label{fig:SchwarzschildPCDiagram}
\end{figure}
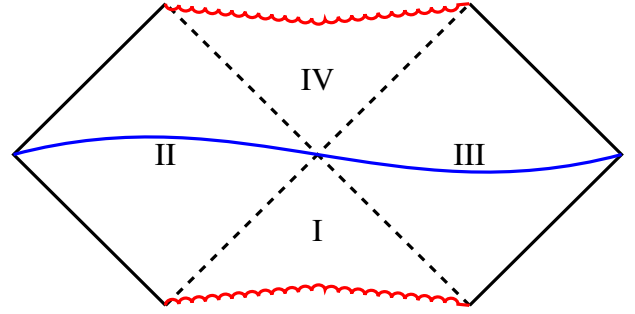

The focus of this letter is Penrose's singularity theorem. A nice formulation is given in \citep{singularitytheorem}:
\begin{theorem}
    Let $(M,g)$ be a globally hyperbolic space-time with non-compact Cauchy surface $\Sigma$. Assume:
    \begin{enumerate}
        \item $R_{\mu\nu}\Dot{\gamma}^\mu\Dot{\gamma}^\nu\geq0$ along all null geodesics $\gamma$;
        \item $M$ contains a future trapped surface.
    \end{enumerate}
    Then $(M,g)$ has incomplete future-directed null geodesics.
    \label{thm:SingThm}
\end{theorem}
This theorem begins with a condition on the geometry of the space-time -- namely, that it is globally hyperbolic. Simply put, this condition requires that all null or timelike geodesics must pass through a spacelike Cauchy hypersurface in the space-time \citep{globalhyperbolicity}. In the case of the Schwarzschild black hole, a Penrose-Carter diagram clearly indicates the existence of a Cauchy hypersurface, as seen in Fig.~\ref{fig:SchwarzschildPCDiagram}.
The next requirement for the theorem is $R_{\mu\nu}\Dot{\gamma}^\mu\Dot{\gamma}^\nu\geq0$. Although this is a condition on the Ricci curvature, this implies a condition of the energy and momentum of the space-time through the Einstein equations. The condition essentially states that the energy density of a region of space cannot be negative. Finally, the space-time must contain a future trapped surface -- a surface from which nothing inside can escape to the outside. All of these conditions together imply that the space-time contains incomplete future-directed null geodesics. To understand these conditions in more detail, a few definitions are needed for a space-time $(M,g)$:
\begin{definition}[Geodesics]
    \label{def:geodesic}
    A \textit{geodesic} $\gamma:(a,b)\rightarrow M$, on a manifold $M$ where $a<0<b$, is a solution to the geodesic equation
    \begin{equation*}
        \Ddot{\gamma}^\mu=-{\Gamma^\mu}_{\alpha\beta}\Dot{\gamma}^\alpha\Dot{\gamma}^\beta, \label{eqn:geodesic}
    \end{equation*}
    with given initial values $\gamma(0)$ and $\Dot{\gamma}(0)$. Here $\gamma$ is parameterised with affine parameter $\lambda$.
\end{definition}
\begin{definition}[Complete geodesics]
    \label{def:completegeo}
    A \textit{future complete geodesic} is as given in Def.~\ref{def:geodesic}, where $b=\infty$. Similarly, a \textit{past complete geodesic} has $a=-\infty$.
\end{definition}
\begin{definition}[Inextendible geodesics]
    A geodesic is \textit{future inextendible} if its future limit, $\displaystyle\lim_{\lambda\rightarrow b^-}\gamma(\lambda)$, does not exist in $M$. Similarly, it is \textit{past inextendible} if its past limit $\displaystyle\lim_{\lambda\rightarrow a^+}{\gamma(\lambda)}$ does not exist in $M$.
\end{definition}
\begin{definition}[Incomplete geodesics]
    A geodesic is \textit{future incomplete} if and only if it is future inextendible and has $b<\infty$. It is \textit{past incomplete} if and only if it is past inextendible with $-\infty<a$. It is \textit{incomplete} if it is future or past incomplete, or both.
\end{definition}
\begin{definition}[Trapped surface]
A \textit{trapped surface} is a compact, spacelike, two-dimensional submanifold with the property that outgoing future directed light rays converge -- in both directions orthogonal to the submanifold -- everywhere on the submanifold.
\end{definition}
To further illustrate the concept of a future incomplete geodesic, consider a car driving along a road of finite length. Its position in three-dimensional space can be described using time as a parameter, so that its path is given by $\gamma(t)=(x(t),y(t),z(t))$. If this path were \textit{incomplete}, it would mean that for some finite value of $t$, the path would abruptly end, with no way to continue the car’s motion beyond this point. Attempting to extend $t$ further would place the car beyond where the road physically exists. In this analogy, the car should be thought of as a future-directed null geodesic and the road should be thought of as the space-time.

It is commonly assumed by those not familiar with the singularity theorems that the presence of incomplete future-directed null geodesics indicates the presence of a curvature singularity in the space-time. This is fair enough -- they are called the \emph{singularity} theorems! However, the next section will give a concrete example as to how this is not always the case, by applying Penrose's singularity theorem to the Kerr space-time.

\subsection{Implications}

Now that a concrete statement of Penrose's singularity theorem has been put forward, the short discussion in the introduction can be made precise. Namely, what does Penrose's singularity theorem say about the Kerr space-time and its most famous extension beyond the inner horizon?

\begin{figure}[h!]
    \begin{center}
      \begin{tikzpicture}[very thick, decoration = {bent,amplitude=-5}, scale=0.8]
        \draw[dashed] (2,-2) -- (0,0) -- (2,2);
        \draw (2,2) -- (4,0) -- (2,-2);
        \draw[dashed] (-2,-2) -- (0,0) -- (-2,2);
        \draw (-2,-2) -- (-4,0) -- (-2,2);

        \draw[blue] (-4,0) to[out=10, in=190] (4,0);

        \draw (-2,-2) -- (0,-4) -- (2,-2);
        \draw[green] (-2,2) -- (0,4) -- (2,2);

        \draw[red,decoration={bumps,amplitude=-2},decorate] (-2,2) -- (-2, 5);
        \draw[red,decoration={bumps,amplitude=-2},decorate] (2,2) -- (2, 5);

        \draw[fill=black] (-1,3.8) circle (2pt);
        \node[above right] at (-1,3.8) {\( p \)};

        \draw[magenta, dotted] (-1,3.8) -- (-2, 2.8);
        \draw[magenta, dotted] (-1,3.8) -- (3.45, -0.55);

        \node at (0,-2) {\large I};
        \node at (-2,-0.8) {\large II};
        \node at (2,-0.8) {\large III};
        \node at (0,2) {\large IV};
        \node at (0, 5) {\large V};

        \node at (1.4,3) {\( r_- \)};
        \node at (-1.35,3) {\( r_- \)};
        \node at (0.75,1.2) {\( r_+ \)};
      \end{tikzpicture}
    \end{center}
    \caption{A piece of a Penrose-Carter diagram of the maximal analytic extension of the Kerr black hole for $\theta=\pi/2$, extended beyond the inner Cauchy horizon in region IV. The future inner Cauchy horizon is drawn in green, timelike singularities in red and a spacelike hypersurface drawn in blue across the horizontal. The boundary of the domain of dependence of a point $p$ inside the inner Cauchy horizon is drawn in magenta.}
    \label{fig:KerrMaxExtPCDiagram}
\end{figure}
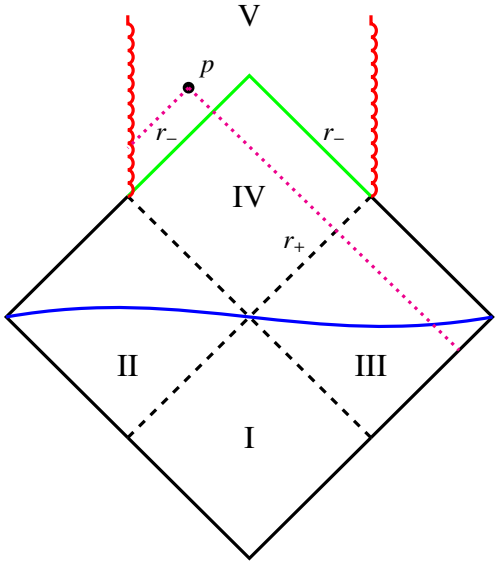

First, considering the maximal analytic extension -- the unique analytic extension in vacuum -- it is evident from Fig.~\ref{fig:KerrMaxExtPCDiagram} that a spacelike hypersurface analogous to a Cauchy surface in Schwarzschild space-time, shown in Fig.~\ref{fig:SchwarzschildPCDiagram}, does not exist in this case. A counterexample is easy to see by way of a point $p$ within $r_-$. It is clear that there exist null and timelike geodesics through this point that never pass through the spacelike curve drawn in Fig.~\ref{fig:KerrMaxExtPCDiagram} and would instead approach the timelike singularity. Hence, the space-time is not globally hyperbolic and so Thm.~\ref{thm:SingThm} does not apply. Thus, in the case of the maximal analytic extension of the Kerr space-time, or in fact \emph{any} extension through $r_-$, Penrose's singularity theorem does not say anything.

\begin{figure}[h!]
    \begin{center}
      \begin{tikzpicture}[very thick, decoration = {bent,amplitude=-5},scale=0.8]
        \draw[dashed] (2,-2) -- (0,0) -- (2,2);
        \draw (2,2) -- (4,0) -- (2,-2);
        \draw[dashed] (-2,-2) -- (0,0) -- (-2,2);
        \draw (-2,-2) -- (-4,0) -- (-2,2);

        \draw[blue] (-4,0) to[out=10, in=190] (4,0);

        \draw (-2,-2) -- (0,-4) -- (2,-2);
        \draw[green] (-2,2) -- (0,4) -- (2,2);

        \node at (0,-2) {\large I};
        \node at (-2,-0.8) {\large II};
        \node at (2,-0.8) {\large III};
        \node at (0,2) {\large IV};
        \node at (0, 5) {\large V};

        \node at (1.4,3) {\( r_- \)};
        \node at (-1.35,3) {\( r_- \)};
        \node at (0.75,1.2) {\( r_+ \)};
      \end{tikzpicture}
    \end{center}
    \caption{A Penrose-Carter diagram of a Kerr black hole for $\theta=\pi/2$ up to the future inner horizon drawn in green, with a Cauchy surface drawn in blue across the horizontal.}
    \label{fig:KerrPCDiagram}
\end{figure}
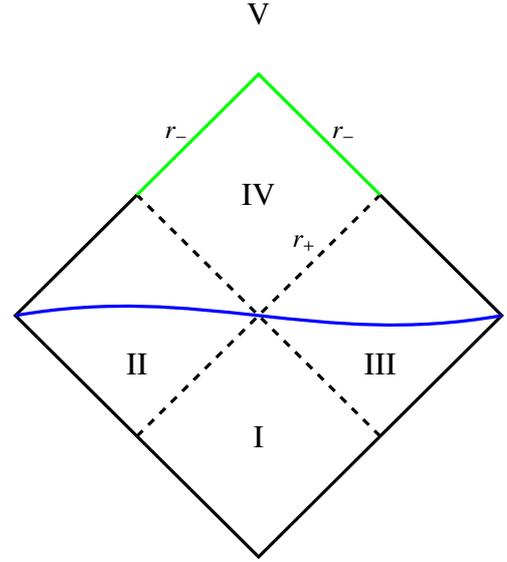

It is important to remark here that the general consensus is that one would expect a different interior solution (the space-time inside the outer horizon $r_+$) to be created in realistic gravitational collapse \citep{Chou_2025}. A different interior would not necessarily contain a Cauchy horizon or a physical curvature singularity if it did not satisfy the assumptions of Thm.~\ref{thm:SingThm}, such as the energy condition \cite{lan2023regular}. Further, if one considers the case given in Fig.~\ref{fig:KerrPCDiagram} as the interior solution for $r \geq r_-$, no matter how one extends beyond the inner Cauchy horizon, it will still be a Cauchy horizon and the assumptions of Thm.~\ref{thm:SingThm} will remain unsatisfied in the extended space-time.

Now consider the space-time only defined for $r \geq r_-$. Fig.~\ref{fig:KerrPCDiagram} presents a Penrose-Carter diagram for this space-time. It is clear from this picture that there exists a non-compact Cauchy hypersurface and so the space-time is globally hyperbolic, satisfying the first assumption of Thm.~\ref{thm:SingThm}. 
So, if one can show that the two remaining conditions of Thm.~\ref{thm:SingThm} hold, then Penrose's theorem tells us that the space-time will have incomplete future-directed null geodesics.

\section{Application to the Kerr space-time}

\label{sec:AppToKerr}

\subsection{Satisfying Penrose's theorem}
\label{sec:SatisfyPenrose}

This section demonstrates, through a concrete example, that Thm.~\ref{thm:SingThm} holds for the Kerr space-time defined up to the inner Cauchy horizon $r_-$. This is most easily demonstrated with an explicitly worked example using the space-time expressed both in Kerr-Schild coordinates (as in Eq.~\eqref{eq:KerrLE}), as well as in Boyer-Lindquist coordinates \citep{boyer1967maximal, visser2008kerrspacetimebriefintroduction}
\begin{align}
    \text{d}s^2 
    &= \left(1 - \frac{2Mr}{\Sigma}\right)\text{d}t^2
    + \frac{\Sigma}{\Delta}\text{d}r^2
    + \Sigma\text{d}\theta^2
    - \frac{4Mra\sin^2\theta}{\Sigma}\text{d}t\text{d}\phi  \nonumber \\
    &\quad+\left(r^2 + a^2 + \frac{2Mra^2\sin^2\theta}{\Sigma}\right)\sin^2{\theta}\text{d}\phi^2, \label{eq:KerrLE}
\end{align}
where
\begin{equation}
    a = \frac{J}{M}, \qquad
    \Sigma = r^2 + a^2\cos^2{\theta}, \qquad
    \Delta = r^2 - 2Mr + a^2. \label{eq:KerrAuxDefs}
\end{equation}

First, it must be shown that $R_{\mu\nu}\Dot{\gamma}^\mu\Dot{\gamma}^\nu\geq0$ along all null geodesics $\gamma$. This is trivially satisfied, as the Kerr space-time is a vacuum solution with vanishing cosmological constant -- thus $R_{\mu\nu}=0$.

Next, it must be shown that the space-time $(M,g)$ contains a future trapped surface. A given 2-surface $\mathcal{S}$ can be identified as being trapped or not by using an adapted null tetrad  $(l^a,n^a,m^a,\Bar{m}^a)$ where $m^a$ and $\Bar{m}^a$ are tangent to $\mathcal{S}$ and $n^a$ and $l^a$ are orthogonal to $\mathcal{S}$. The convergences $\rho$ and $\rho'$ are then able to be calculated through
\begin{align*}
    \rho&:= m^a\Bar{m}^b\nabla_b l_a, \\
    \rho'&:= \Bar{m}^am^b\nabla_b n_a 
\end{align*}
respectively, where $\nabla$ is the Levi-Civita connection \citep{penrose1986spinors}. The real part of $\rho$ represents the contraction or expansion of a bundle of light rays in the $l^a$ direction when this is positive or negative respectively, with the imaginary part representing the twist in the same direction. The same is true for $\rho'$ but in the $n^a$ direction. Using these geometric scalars computed in an $\mathcal{S}$-adapted null tetrad, condition $2$ of Thm.~\eqref{thm:SingThm} can be written as
\begin{equation}\label{eq:trappedsurfacedefn}
    \text{Re}(\rho) > 0, \quad \text{Re}(\rho') > 0,
\end{equation}
indicating that a bundle of null rays pointing in either the $l^a$ or $n^a$ direction are converging. Of note is that the definition of a trapped surface in terms of the scalars $\rho$ and $\rho'$ can also be used to define the Marginally Outermost Trapped Surface (MOTS) when $\text{Re}(\rho') > 0$ as above but now with $\text{Re}(\rho) = 0$ \citep{thornburg2007event}, assuming that $l^a$ is directed outwards. If this is satisfied on some compact spacelike, 2-surface $\mathcal{S}$ in the $\mathcal{S}$-adapted null tetrad, this indicates that a bundle of null rays in the future outward radial direction $l^a$ do not diverge.

To show that these conditions hold for the Kerr metric, $\mathcal{S}$ must be specified. For this, we use the space-time written in Boyer-Lindquist coordinates. Given that the two roots of $\Delta=0$ yield the inner and outer event horizons $r_-$ and $r_+$, any $\mathcal{S}$ defined by $r=$\;constant chosen between these surfaces will suffice. 

A particularly simple choice of null tetrad can be written in the holonomic (coordinate) basis as
\begin{align*}
    l^\mu &= 
    \frac{1}{2r^5}\Big{\{}\Pi\left[a^2\Delta\sin^2\theta - (a^2 + r^2)^2\right],\;
    \Delta\Pi\Phi\sin\theta,  \\
    &\quad0,\;
    -2 r a M \Pi\left\},\right. \\
    n^\mu &= 
    2r^5\Big{\{}-\frac{1}{\Delta\Pi^3},  \\
    &\quad-\frac{2\Phi\sin\theta}{\Pi^3\left[a^4 + 2r^4 + a^2r(2M + 3r) + a^2\Delta\cos(2\theta)\right]},
    0,  \\
    &\left.\quad\frac{2a M r}{\Delta\Pi^3\left[a^2\Delta - (a^2 + r^2)^2\csc\theta\right]\sin^2\theta}\right\}, \\
    m^\mu &= 
    \frac{1}{\sqrt{2}}\left\{(0,0,
    \frac{1}{\sqrt{\Sigma}},
    \frac{i\sqrt{\Sigma}}{\sin{\theta}\sqrt{(a^2 + r^2)^2 - a^2\Delta\sin^2\theta}}\right\},
\end{align*}
where 
\begin{equation*}
    \Pi := \sqrt{a^2 + 2r^2 + a^2\cos(2\theta)}, \quad
    \Phi := \sqrt{(a^2 + r^2)^2\csc^2\theta - a^2\Delta}.
\end{equation*}
One needs to be careful when defining an adapted null tetrad here, as the meaning of \emph{future} is represented differently whether one is inside or outside $r_+$. Namely, it is clear that \emph{outside} $r_+$, the notion of going into the future is represented by increasing $t$. However, when \emph{inside} $r_+$, it is represented by decreasing $r$ as one is inescapably drawn further into the black hole. This oddity is due to the change in the causal structure when passing through the outer horizon. The above tetrad is indeed future pointing inside $r_+$ as $n^0,\,n^1 < 0$ and $l^0 > 0,\,l^1 < 0$ for at least a large class of choices of $a$ and $M$, and certainly the choice eventually chosen here.

Given the adapted null tetrad, the associated convergences $\rho$ and $\rho'$ are found to be
\begin{align}
    \begin{split}
        \rho
        &= \frac{4r^6\csc\theta(a^2 + r^2) + 2a^2r^5\sin\theta(M - r)}
        {\Pi^3\Phi\left[a^4 + 2r^4 + a^2r(2M + 3r) + a^2\Delta\cos(2\theta)\right]},
        \label{eq:rho}
    \end{split} \\
    \begin{split}
        \rho'
        &= -\frac{\Delta\Pi^5\left[4r^3 + a^2(M + 3r) + a^2\cos(2\theta)(r - M)\right]}
        {64r^5\Sigma^2\Phi^3\sin^3\theta}
        \\
        &\quad\times \left[a^4 + 2r^4 + a^2r(2M + 3r) + a^2\Delta\cos(2\theta)\right].
        \label{eq:rhop}
    \end{split}
\end{align}
To obtain explicit values for these expressions, the choice $2a=M=1$ is now taken. With this choice, remembering that the event horizons are defined by the solutions to $1/g_{rr}=0$, they obtain the values $r_- \approx 0.13398$ and $r_+ \approx 1.86603$. Then the 2-surface $\mathcal{S}$ is taken to be the surface defined by $r=\sqrt{3}/2$ and some arbitrary choice of $t$.

\begin{figure}
    \centering
    \includegraphics[width=0.75\linewidth]{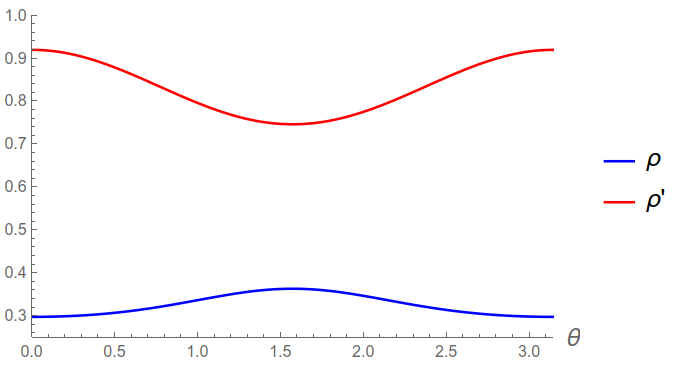}
    \caption{The convergences $\rho,\,\rho'$ for outgoing and ingoing rays respectively, plotted against $\theta$ for a choice of $a=1/2$, $M=1$ in a null tetrad adapted to the 2-surface $\mathcal{S}$.}
    \label{fig:convergences}
\end{figure}

Evaluating $\rho$ and $\rho'$, as given by Eqs~\eqref{eq:rho} and \eqref{eq:rhop}, on $\mathcal{S}$, the null rays converge in both future orthogonal directions to $\mathcal{S}$ (since they are both positive and real), as seen in Fig.~\ref{fig:convergences}. Hence Eq.~\eqref{eq:trappedsurfacedefn} is satisfied and $\mathcal{S}$ is a future trapped surface.

\subsection{Incomplete future-directed null geodesics}
At this point, all the assumptions in Thm.~\ref{thm:SingThm} have been shown to be satisfied for our space-time and choice of 2-surface $\mathcal{S}$, so it remains to calculate the conclusion of the theorem -- the incomplete future-directed null geodesics. Although exact expressions exist for these in terms of elliptic functions \citep{gralla2020null}, they are complicated and for exemplary purposes, a simple numerical method will be used here.

To achieve this, the Kerr space-time is considered in Kerr-Schild coordinates Eq.~\eqref{eq:KerrLE}, due to the coordinate singularity at $r_-$ when using Boyer-Lindquist coordinates. Geodesics lying within the equatorial plane $z=0$ are considered for simplicity. These are governed by the Euler-Lagrange equations 
\begin{equation*}
    \frac{d}{d\lambda}\left(\frac{d\mathcal{L}}{d\Dot{x}^\mu}\right)-\frac{d\mathcal{L}}{dx^\mu}=0,
\end{equation*}
with Lagrangian
\begin{equation*}
    \mathcal{L}=\frac12g_{\mu\nu}\Dot{x}^\mu\Dot{x}^\nu,
\end{equation*}
where the geodesic $\gamma^\mu = x^\mu = x^\mu(\lambda)$ is parametrized by an affine parameter $\lambda$ and described by coordinates $x^\mu = \{t(\lambda),\,x(\lambda),\,y(\lambda),\,0\}$.
Supplementing these equations is the condition that $\Dot{x}^\mu$ be null
\begin{equation}
    0 
    = g_{\mu\nu}\dot{\gamma}^\mu\dot{\gamma}^\nu. \label{eq:nullgeocond}
\end{equation}
It can be easily checked that the vector
\begin{equation}
    N^\mu 
    = \left\{1,\,
    -\frac{r x + a y}{a^2 + r^2},\,
    -\frac{r y - a x}{a^2 + r^2},\,
    0
    \right\}
\end{equation}
is null. Evaluated at the point $\{0,\,1,\,0,\,0\}$ (where $r=\sqrt{3}/2$ and so this point lies on $\mathcal{S}$), $N^\mu$ is $\{1,-\sqrt{3}/2\,,1/2,\,0\}$, and is thus inward pointing.

The three Euler-Lagrange equations (where the fourth is trivially satisfied with $z(\lambda)=0$) with these initial conditions are then solved numerically in \textsc{Mathematica} with $\texttt{NDSolve}$. The numerical implementation was confirmed by showing convergence at the correct order for decreasing step size.

\begin{figure}[hbt!]
    \centering
    \begin{subfigure}{.45\linewidth}
      \includegraphics[width=\linewidth]{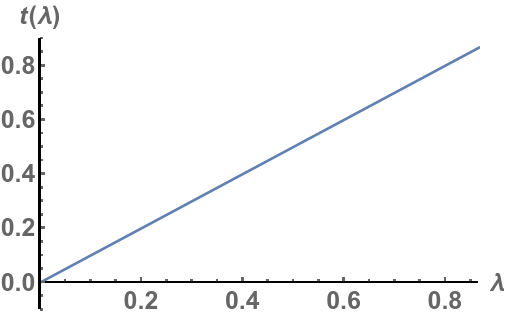}
      \caption{$t(\lambda)$}
    \end{subfigure}\hfill
    \begin{subfigure}{.45\linewidth}
      \includegraphics[width=\linewidth]{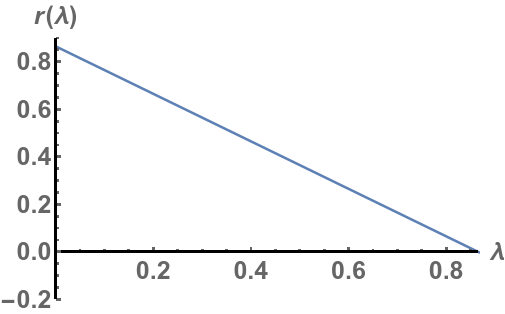}
      \caption{$r(\lambda)$}
    \end{subfigure}
    \caption{The position of the ingoing null geodesic as a function of $\lambda$.}
    \label{fig:nullgeos}
\end{figure}

The results of the ray tracing for $t(\lambda)$ and $r(\lambda)$ (where $r(\lambda)$ is computed from the solutions of $x(\lambda)$ and $y(\lambda)$) are shown in Fig.~\ref{fig:nullgeos}.
The numerical methods work up until the singularity at $r=0$ before stopping. It is clear that this null geodesic passed through $r_-$ without any unusual behavior. In particular, it did so at a \emph{finite affine time}.\footnote{The fact that $\lambda$ is an affine parameter follows as one can show that the Euler-Lagrange equations used with our choice of Lagrangian is equivalent to the affinely parametrized geodesic equation.} Thus, if the space-time was taken to be defined only for $r\geq r_-$, the conclusion of Thm.\ref{thm:SingThm} is verified explicitly.

It is now of interest to check whether there is a physical curvature singularity where the geodesic stopped in the sense of an unbound scalar curvature invariant. The Kretschmann scalar is more easily given in Boyer-Lindquist coordinates, and is
\begin{align*}
    I
    &= R_{\alpha\beta\gamma\delta}R^{\alpha\beta\gamma\delta} \\
    &= \frac{48M^2(r^6-15a^2r^4\cos^2{\theta}+15a^4r^2\cos^4{\theta}-a^6\cos^6{\theta})}{(r^2+a^2\cos^2{\theta})^6}
\end{align*}
which simplifies nicely for our choice of $2a=M=1$ to $48/r^6$ on $z=0$ ($\theta=\pi/2$), clearly finite everywhere but $r=0$.

\section{Discussion}\label{sec:discussion}
A detailed worked example has been provided, applying Penrose's singularity theorem to the Kerr space-time in its unextended form (defined up to the inner Cauchy horizon) and an explanation given as to why the theorem is not valid for any space-time that extends through this horizon.

First, looking at the maximal analytic extension, it is clear in this case that the space-time is not globally hyperbolic as there does not exist a Cauchy surface (see Fig.~\ref{fig:KerrMaxExtPCDiagram}) and so Penrose's singularity theorem does not apply. It is known however, that there are timelike physical curvature singularities in this extended space-time, as shown at the end of Sec. \ref{sec:AppToKerr}.

Now consider the non-extended space-time. In this case, the space-time is taken to be globally hyperbolic, and can only be defined up to at most the inner Cauchy horizon. Hence, when future-directed null geodesics end there at finite values of their affine parameter, they are inextendible and incomplete. Penrose's singularity theorem holds for this case as the requirements are met, as given in detail in the preceding section. Now consider the case where the space-time is cut off at some $r_0$ lying within $r_\pm$. Penrose's theorem holds as in the above case, but the important point to make here is that the existence of future-directed incomplete null geodesics have identified that the space-time \emph{can be extended} rather than identifying a physical curvature singularity in the sense of a diverging scalar curvature invariant. In the maximal case $r_0=r_-$, the Cauchy horizon signals that the space-time may be extended, but in such a way that the domain of dependence of points in the extension require more information than is contained in the non-extended space-time. In other words, this means that the space-time will lose global hyperbolicity and Thm.~\ref{thm:SingThm} will no longer hold. In the maximal analytic extension, boundary data on the timelike singularities would be required, see Fig.~\ref{fig:KerrMaxExtPCDiagram}.

Although the inner Cauchy horizon is not a physical curvature singularity in the sense of an unbound Kretschmann scalar, it is still a singularity \emph{in some sense}. The inner Cauchy horizon is thought to be unstable under small perturbations. This has been investigated for example by Simpson and Penrose for the Reissner-Nordstr\"{o}m (charged, static, spherically symmetric) black hole\footnote{While this black hole solution shares a similar conformal structure with Kerr and benefits from the simplifying assumption of spherical symmetry, the Kerr space-time remains the more physically relevant example and thus was chosen as the primary focus of the paper.} \citep{reissnernordstromunstableperturbation}. They found that asymmetric perturbations of electromagnetic fields on a Reissner-Nordstr\"{o}m background became unstable on the inner Cauchy horizon and argue this should carry over to the fully coupled case. Poisson and Israel performed a similar analysis, and claim that this result may be extended to the Cauchy horizon of Kerr \citep{PoissonIsraelCauchySingularity}. Hence, in a realistic gravitational collapse scenario, it may very well be the case that the inner Cauchy horizon present in the Kerr space-time is replaced by a physical curvature singularity. 

\section*{Acknowledgements}
The authors thank David Wiltshire for recommending this work to the special issue's guest editors and Marco Galoppo and J\"{o}rg Hennig for giving feedback on the manuscript.

\end{document}